\documentclass[manuscript]{aastex62}

\usepackage{txfonts}
\usepackage{latexsym,bm}
\usepackage{multirow}

\submitjournal{ApJ}

\shorttitle{Sunspot Number Model}
\shortauthors{Qin and Wu}

\begin{document}

\title{A Model of Sunspot Number with Modified Logistic Function}

\correspondingauthor{G. Qin}
\email{qingang@hit.edu.cn}

\author[0000-0002-3437-3716]{G. Qin}
\affiliation{School of Science, Harbin Institute of Technology, Shenzhen, 
518055, China; qingang@hit.edu.cn}

\author[0000-0002-5776-455X]{S.-S. Wu}
\affiliation{School of Science, Harbin Institute of Technology, Shenzhen, 
518055, China; qingang@hit.edu.cn}
\affiliation{State Key Laboratory of Space Weather, National
Space Science Center, Chinese Academy of Sciences,
Beijing 100190, China}
\affiliation{College of Earth Sciences, University of
Chinese Academy of Sciences, Beijing 100049, China}

\begin{abstract}
Solar cycles are studied with the Version 2 monthly smoothed 
international sunspot number, the variations of which are found to be
well represented by the modified logistic differential equation with 
four parameters: maximum cumulative sunspot number or total sunspot 
number $x_m$, initial cumulative sunspot number $x_0$, maximum emergence 
rate $r_0$, and asymmetry $\alpha$. A two-parameter function is
obtained by taking $\alpha$ and $r_0$ as fixed value. In addition, it 
is found that $x_m$ and $x_0$ can be well determined at the start of a 
cycle. Therefore, a prediction model of sunspot number is 
established based on the two-parameter function. The prediction for 
cycles $4-23$ shows that the solar maximum can be predicted with 
average relative error being 8.8\% and maximum relative 
error being 22\% in cycle 15 at the start of solar cycles if 
solar minima are already known. The quasi-online method for determining 
solar minimum moment shows that we can obtain the solar minimum 14 months 
after the start of a cycle. Besides, our model can predict the cycle 
length with the average relative error being 9.5\% 
and maximum relative error being 22\% in cycle 4. Furthermore, we 
predict the sunspot number variations of cycle 24 with the relative 
errors of the solar maximum and ascent time being 1.4\% and 
12\%, respectively, and the predicted cycle length is 
11.0 (95\% confidence interval is 8.3$-$12.9) years. The comparison 
to the observation of cycle 24 shows that our prediction model has 
good effectiveness.
 
\end{abstract}


\section{Introduction}

Sunspot number is a good indicator of solar activity and its change 
follows an 11-year cycle. Solar activity has influence on the galactic 
cosmic ray intensity \citep[e.g.,][]{McDonald98, QinAShen17, ShenAQin18} 
that has impact on the health of astronauts and safety of spacecraft. It 
is also suggested that solar activity may affect Earth's climate 
\citep[e.g.,][]{Haigh2007}. There is evidence that the energy spectrum 
of ground-level enhancement (GLE) of solar energetic particle events is 
relevant with the 10.7~cm solar radio flux \citep{WuEA2018} that is 
highly correlated with sunspot number \citep[e.g.,][]{HollandEA1984, 
Hathaway2015}, so that the prediction of the time profile of sunspot 
number is helpful for estimating the spectra of potential GLEs. In 
addition, the fact that sunspot number is related to other solar 
parameters may guide us to better understand the origin of solar cycles.

In order to describe the variations of sunspot number in one solar cycle, 
\citet{StewartEA1938} adopted Pearson's Type III distribution function 
with a power law for the rising phase and an exponential for the decline 
phase. Furthermore, \citet{HathawayEA1994} constructed a quasi-Planck 
function similar to that of \citet{StewartEA1938} but with a fixed 
(cubic) power law for the rising phase and a Gaussian for the decline 
phase, which is given by \citep{HathawayEA1994, Hathaway2015}:
\begin{equation}
f\left(t\right)=a\left(t-t_0\right)^3
 \left[\exp{\left(\frac{t-t_0}{b}\right)^2}-c\right]^{-1}
 \label{eq:quasi_Planck}
\end{equation}
with four parameters, namely, amplitude $a$, starting time $t_0$, rise 
time $b$, and asymmetry $c$. They found that the asymmetry $c$ can be 
taken as a fixed value and the rise time $b$ is relevant with the 
amplitude $a$, so that the function could be reduced to a two-parameter 
form and the sunspot number could still be fitted well. They found that 
the amplitude can be estimated at the start of a solar cycle by using 
the correlation between the amplitude and the length of the previous 
cycle, and the relative error is within about 30\%. At some time 
$\Delta t$ that a solar cycle has progressed one can use this model to 
fit the observed data and thus one can predict the sunspot number for 
the rest of the cycle. They concluded that the accuracy of the predicted 
amplitude is within about $20$\% and $10$\% if $\Delta t=30$ and $42$ 
months, respectively. \citet{Volobuev2009} suggested a function
that is similar to Maxwell distribution with three parameters to 
represent the sunspot number \citep[e.g.,][]{RoshchinaEA2011}, 
and they also obtained a two-parameter form by reducing one of the 
parameters. Note that, they considered it to be one-parameter fit 
because the parameter of starting time was neglected to fit. They 
found that the fitting effects of their model were similar to that of 
\citet{HathawayEA1994}. What's more, they concluded that their 
empirical model did better than dynamo ones in fitting the sunspot 
number. By introducing an asymmetry factor $\alpha$ to describe the 
asymmetry of the sunspot number during a solar cycle, \citet{Du2011} 
introduced a modified Gaussian function
\begin{equation}
f\left(t\right)=A\exp{\left(\frac{-\left(t-t_m\right)^2}
 {2B^2\left[1+\alpha\left(t-t_m\right)\right]^2}\right)}
\label{eq:mod_Gaussian}
\end{equation}
with four parameters, among which they found that $B$ and $\alpha$ can 
be obtained by quadratic functions of $t_m$, and thus the number of 
parameters is reduced to two. Based on the two-parameter function, they 
concluded that the accuracy of predicted maximum sunspot number is 
within $15$\% if $\Delta t=25$ months.

\citet{Gnevyshev1963} found that during solar maximum sunspot number 
usually has double-peak with a gap in between, which is called Gnevyshev 
gap. In order to fit this phenomenon \citet{SabarinathEA2008} suggested 
the modified binary mixture of the Laplace distribution functions
\begin{equation}
f\left(t\right)=\frac{A_1}{2S_1}\exp{\left(\frac{-|t-M_1|}{S_1}\right)}
 +\frac{A_2}{2S_2}\exp{\left(\frac{-|t-M_2|}{S_2}\right)}
\end{equation}
with six parameters as the model of sunspot number. They also obtained 
a two-parameter function by using empirical values for $M_1$, $M_2$, 
$S_1$, and $S_2$. Recently, \citet{LiEA2017} proposed a simplified binary 
mixture of Gaussian functions as follows:
\begin{equation}
f\left(t\right)=A_1\exp{\left(\frac{-\left(t-M_1\right)^2}{S_1}\right)}
 +A_2\exp{\left(\frac{-\left(t-M_2\right)^2}{S_2}\right)}
\end{equation}
with six parameters too, which, however, shows better results in the 
double-peak to fit monthly smoothed sunspot number data. In addition,
they found that the function could be reduced to a three-parameter 
form since the parameters $M_2$, $S_1$, and $A_2$ could be represented 
by $M_1$.

It is shown that empirical functions can be used to represent the 
variations of sunspot number during a solar cycle with demonstrated 
prediction power, but they usually do not obtain a good prediction 
result for a solar cycle until 2 to 3 years that the solar cycle has 
progressed, i.e., $\Delta t=2$ to $3$ years
\citep{HathawayEA1994, Du2011, Hathaway2015}. In this paper we use the 
modified logistic differential equation to reproduce the variations of 
sunspot number during solar cycles, and thus a prediction model is 
established with good effectiveness. In section~\ref{sec:data}, sunspot 
data are presented. In section~\ref{sec:model}, our new sunspot number 
model with modified logistic function is presented. In 
section~\ref{sec:result}, we show the results of the new model. In 
section~\ref{sec:prediction}, the prediction ability of our model is 
presented. And conclusions are presented in section~\ref{sec:conclusion}.

\section{Sunspot data}
\label{sec:data}

\begin{figure}
\epsscale{1} \plotone{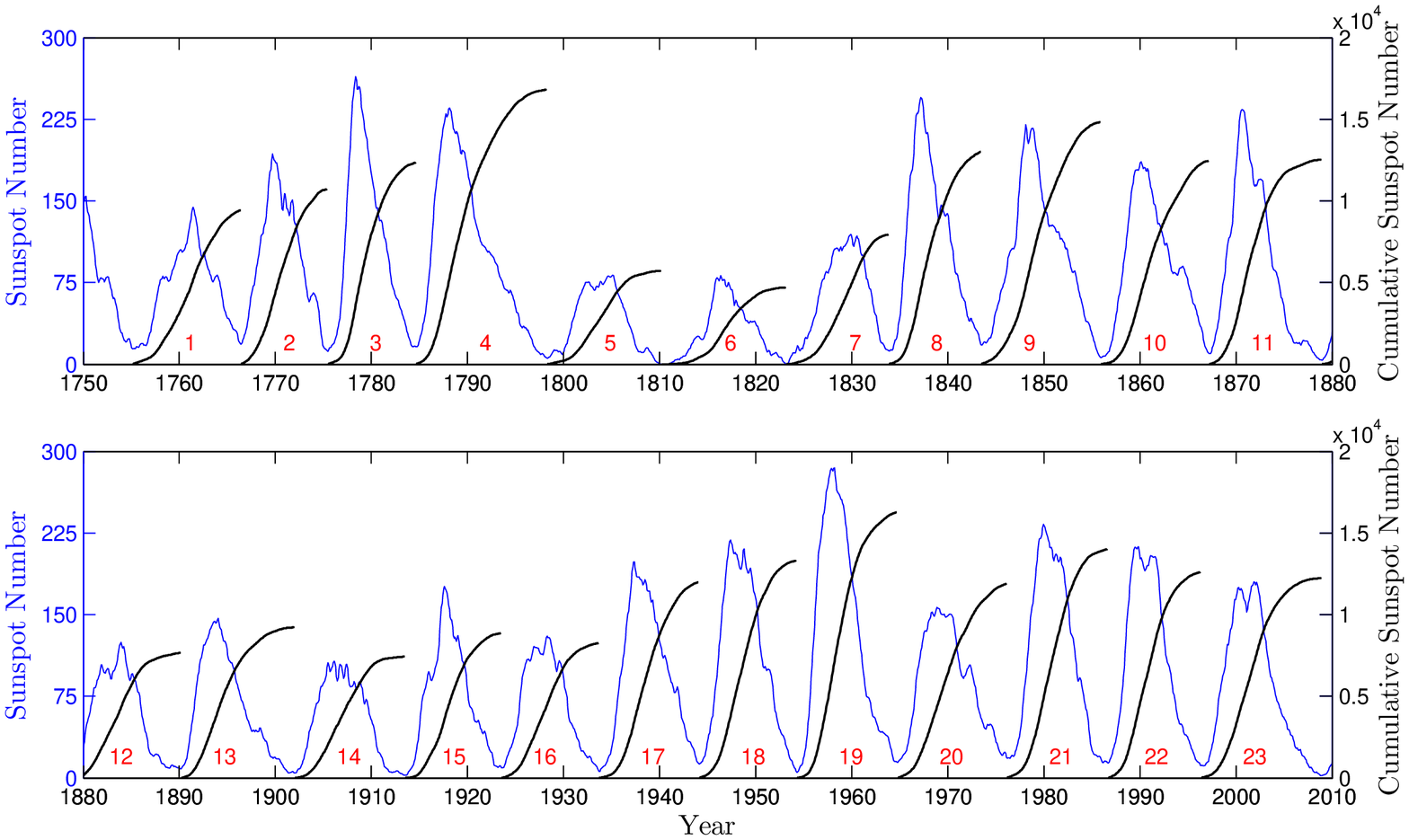}
\caption{Blue and black curves are monthly smoothed sunspot number and
 cumulative sunspot number, respectively, for cycles $1-23$. The numbers 
 in the figure indicate the cycle number.}
\label{fig:solarCycles}
\end{figure}

In this work, we use the Versions 1 and 2 \citep{CletteEA2015} 
monthly and monthly smoothed international sunspot number, which can 
be downloaded from http://www.sidc.be/silso/. Monthly smoothed 
sunspot number, which is obtained by using the standard smoothing with a
13-month running mean centered on the month of interest with half weights
for the months at the start and end \citep[e.g.,][]{Hathaway2015},  
is adopted to study the shape of solar cycles. The solar maximum is 
obtained by taking the mathematical maximum of the monthly smoothed 
sunspot number in a solar cycle, and the solar minimum is obtained by 
taking the mathematical minimum of the monthly smoothed sunspot number 
in the period from the previous solar maximum to current one. Note that, 
if the minimum/maximum value is occurred more than once in a period, 
the first occurrence is taken as the epoch of the minimum/maximum 
\citep[e.g.,][]{Kakad2011}. For example, the minimum value between 
the solar maximum epoch of cycle 22 and that of cycle 23 is 11.2 
that occurred in May 1996 and August 1996, we choose May 1996 as
the solar minimum epoch of solar cycle 23. The blue and black curves 
in Figure~\ref{fig:solarCycles} represent monthly smoothed
sunspot numbers of Version 2 and their cumulative values for 
solar cycles, respectively. In other words, the blue curves are the 
time derivative of the black ones. 
 
\section{Sunspot number model with modified logistic function}
\label{sec:model}

\subsection{Logistic function}

Logistic function is proposed by \citet{Verhulst1838} for modeling 
population growth, and the function has been successfully used in many 
fields such as statistics, machine learning, chemistry, physics, 
economics, and sociology. In the model, the growth of population is 
related to the number of the current population, which can be written as:
\begin{equation}
\frac{dx}{dt}=rx, \label{eq:dxdt}
\end{equation}
where $x$ and $r$ are the current population and growth rate, 
respectively. Due to the limitation of environmental factors, the growth
rate decreases with the increase of population, which has a maximum value
$x_m$. Consequently, the growth rate $r$ can be written as:
\begin{equation}
r(x)=r_0-kx, \label{eq:r_k}
\end{equation}
where $r_0$ and $k$ are constants with $r_0$ being called intrinsic growth
rate. When the population reaches its maximum value $x_m$, the growth rate 
$r$ becomes zero, i.e., $r(x_m)=0$. Therefore, the constant $k$ could be 
eliminated, and Equation~(\ref{eq:r_k}) is rewritten as:
\begin{equation}
r(x)=r_0\left(1-\frac{x}{x_m}\right). \label{eq:r_xm}
\end{equation}
Substituting Equation~(\ref{eq:r_xm}) into Equation~(\ref{eq:dxdt}), the
differential equation of population is given by:
\begin{equation}
\frac{dx}{dt}=r_0\left(1-\frac{x}{x_m}\right)x. \label{eq:x_diff}
\end{equation}
Using the initial condition $x|_{t=0}=x_0$, the integral equation of
population is derived as:
\begin{equation}
x\left(t\right)=x_m\left[1+\left(\frac{x_m}{x_0}-1\right)e^{-r_0t}\right]^{-1},
\label{eq:x_int}
\end{equation}
which is called logistic function.

\begin{figure}
\epsscale{0.6} \plotone{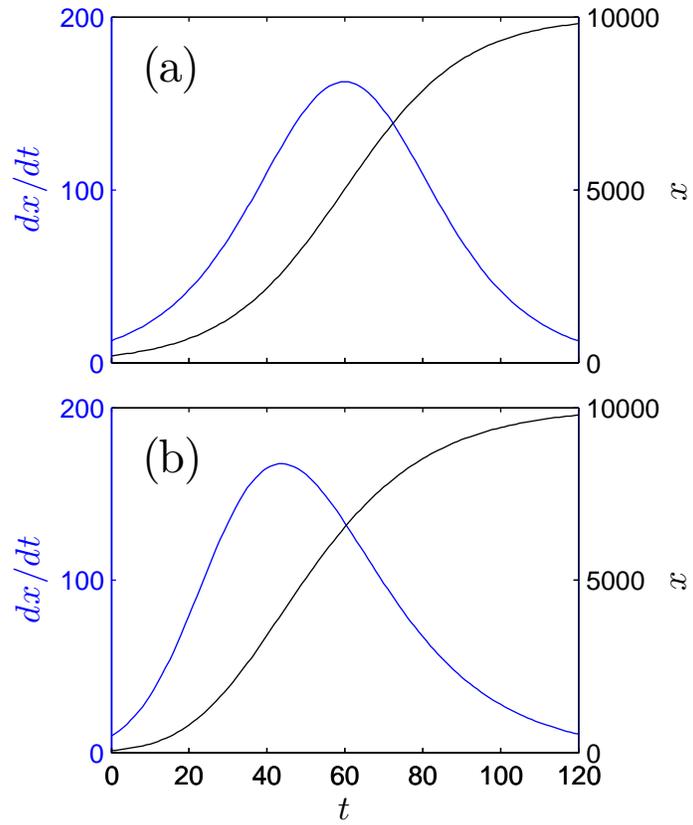}
\caption{(a) Black and blue curves are logistic function and
its derivative, respectively, with 
$x_0$=200, $r_0$=0.065, and $x_m$=10000. (b) Same as Panel (a) but for
modified logistic model with $\alpha=0.2$, $x_0$=60, $r_0$=0.25, 
and $x_m$=10000.}
\label{fig:logisticCurves}
\end{figure}

In Figure~\ref{fig:logisticCurves}(a), the black curve shows the logistic 
function, and the blue curve is the derivative of the black one. It can
be seen that the black curve in Figure~\ref{fig:logisticCurves}(a) 
(logistic function) is similar to the black curve in 
Figure~\ref{fig:solarCycles} (cumulative sunspot numbers), so that one
may use the logistic model to study the variations of solar cycles. 
However, the shape of logistic differential equation 
(Equation~\ref{eq:x_diff}) in Figure~\ref{fig:logisticCurves}(a) is 
symmetrical about its peak, while the shape of solar cycles is usually 
asymmetrical about its peak. Therefore, the logistic function needs to
be modified to show asymmetry.

\subsection{Modified logistic function}

In order to obtain an asymmetrical logistic differential equation,
we replace $x$ with $x^\alpha$ in Equation~(\ref{eq:r_k}) to get:
\begin{equation}
r(x)=r_0-kx^\alpha,
\label{eq:r_k_m}
\end{equation}
where $\alpha$ is a positive constant. It is noted that in the new 
Equation~(\ref{eq:r_k_m}) the relationship between $r$ and $x$ is no 
longer linear. The modified differential and integral equations 
corresponding to Equations~(\ref{eq:x_diff}) and (\ref{eq:x_int}) are 
derived as:
\begin{eqnarray}
\frac{dx}{dt}&=&r_0\left(1-\frac{x^\alpha}{x_m^\alpha}\right)x,
 \label{eq:x_diff_mod}\\
x\left(t\right)&=&x_m {\left[1+\left(\frac{x_m^\alpha}{x_0^\alpha}-1\right)
 e^{-\alpha r_0t}\right]^{-1/\alpha}}, \label{eq:x_int_mod}
\end{eqnarray}
which is very similar to the generalized logistic function
\citep{Richards1959, Birch1999, Balakrishnan2010}. In 
Figure~\ref{fig:logisticCurves}(b), the black and blue curves show 
the modified logistic function and its differential results with 
$\alpha=0.2$, respectively. We can see that the blue curve shows 
asymmetry about its peak, i.e., it increases rapidly before the peak 
while decreases slowly after the peak.

\subsection{Models of sunspot number}

The Equations~(\ref{eq:x_diff}) and (\ref{eq:x_diff_mod}) could be 
rewritten as follows:
\begin{eqnarray}
f\left(x\right)&=&r_0\frac{x}{x_m}\left(x_m-x\right)=
r_0\frac{x}{x_m}x_r, \label{eq:x_diff_re}\\
f\left(x\right)&=&r_0\frac{x}{x_m}\frac{x_m^\alpha-x^\alpha}{x_m^{\alpha-1}}
=r_0\frac{x}{x_m}x_r^\prime,
 \label{eq:x_diff_mod_re}
\end{eqnarray}
where $dx/dt$ is replaced with $f$ that denotes the sunspot number. The 
parameters $x$ and $x_m$ represent current population and maximum 
population, respectively, in Ecology. However, in this work we call $x$ 
and $x_m$ as the current cumulative sunspot number and maximum 
cumulative sunspot number or total sunspot number of a solar 
cycle. Thus, the term $x_r=x_m-x$ in Equation~(\ref{eq:x_diff_re}) 
indicates the residual sunspot number, which changes from $x_m$ to zero. 
In other words, the remaining sunspot number that the Sun will produce 
in the solar cycle is $x_r=x_m-x$. In the modified 
Equation~(\ref{eq:x_diff_mod_re}), the term $x_r=x_m-x$ in 
Equation~(\ref{eq:x_diff_re}) is modified as 
$x_r^\prime=\left(x_m^\alpha-x^\alpha\right)/x_m^{\alpha-1}$, which also 
changes from $x_m$ to zero. Thus the term $x_r^\prime$ represents the 
modified residual sunspot number. With the definition of $x_r=x_m-x$, the 
term $r_0\left(x/x_m\right)$ could be called as emergence rate of sunspot 
which changes from zero to $r_0$, so that $r_0$ indicates the maximum 
emergence rate of sunspot. In addition, the parameters $x_0$ and $\alpha$ 
can be called initial cumulative sunspot number and asymmetry, 
respectively. Finally, the cumulative sunspot number $x\left(t_e\right)$ 
at the end of solar cycle $t_e$ expressed by Equation (\ref{eq:x_int_mod}) 
is indicated by $x_e$. Note that the variable $t$ in
Equation~(\ref{eq:x_int_mod}) represents the months that a solar cycle
has progressed. Thus, we obtain the model of sunspot number with 
a modified logistic differential equation expressed with 
Equation~(\ref{eq:x_diff_mod_re}).

\section{Results}
\label{sec:result}

\subsection{Fitting results}

\begin{figure}
\epsscale{1} \plotone{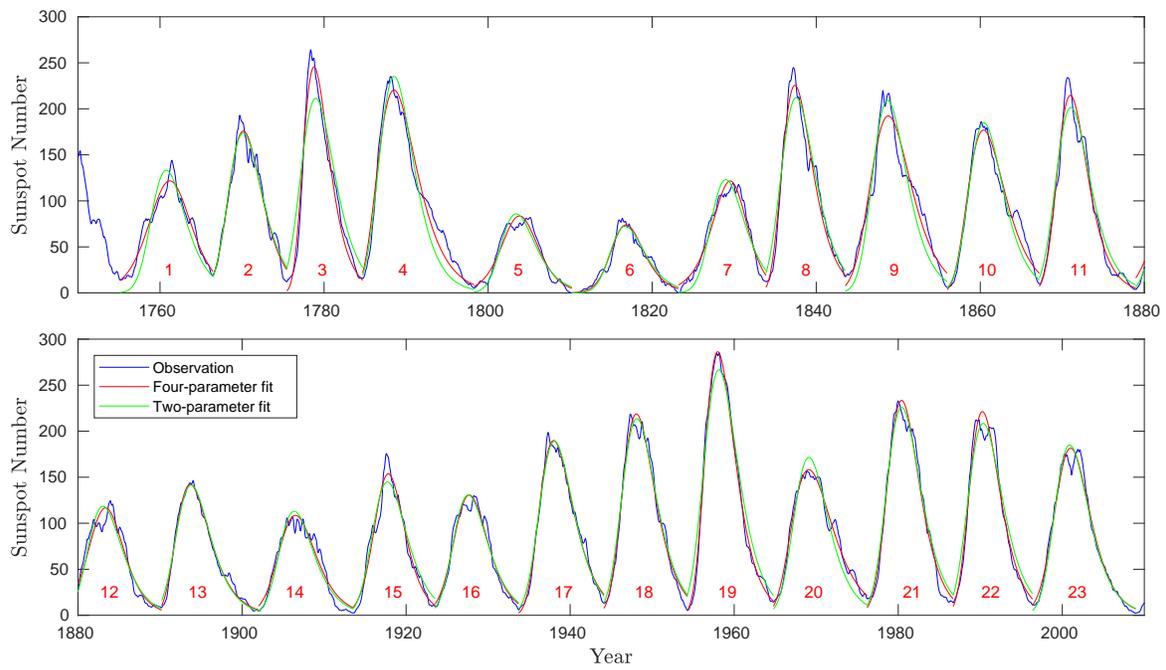}
\caption{Monthly smoothed sunspot number curve (blue) and the fitting 
results by the four-parameter modified logistic differential equation 
(red) and the two-parameter modified logistic differential equation 
(green), for cycles $1-23$.}
\label{fig:solarCycles_fit}
\end{figure}

We fit the Version 2 monthly smoothed sunspot number data with 
the model from the modified logistic differential equation for solar 
cycles from $1$ to $23$, using the non-linear least squares
technique based on BFGS algorithm that is an iterative method for 
solving unconstrained nonlinear optimization problems
\citep{Fletcher1987} to optimize the free parameters. In this work, 
the initial values of $\alpha$ and $r_0$ are all set to 0.2, and that 
of $x_m$ is set to the observed total sunspot numbers of a cycle. 
Besides, the initial value of $x_0$ can be obtained by combining
Equation~(\ref{eq:x_diff_mod_re}) with the conditions $x|_{t=0}=x_0$ 
and $f(x_0)=S_0$, which is the monthly smoothed sunspot number of the
first month of a cycle. Based on the configuration of the initial values,
the fitting shows good stability and convergence can be achieved for
all of the cycles. The red curves in Figure~\ref{fig:solarCycles_fit} 
exhibit the fitting results, and the blue curves are the observations. 
From the figure we can see that the model can fit sunspot number data 
very well. The four fitting parameters $\alpha$, $r_0$,
$x_0$, and $x_m$ with their uncertainties 
and the value $x_e$ are shown in Table \ref{tab:4para}.

\subsection{Evaluating of the fitting results}

In order to evaluate the fitting results, we calculate the fitting 
deviation $\sigma$ \citep{Li1999}
\begin{equation}
\sigma=\sqrt{\frac{\sum_{i=1}^N \left(S_i-f_i\right)^2}{N}},
\label{eq:sigma}
\end{equation}
and the correlation coefficient $CC$ \citep{BevingtonEA2003}
\begin{equation}
CC=\sqrt{1-\frac{\sum_{i=1}^N \left(S_i-f_i\right)^2}
 {\sum_{i=1}^N \left(S_i-\overline{S}\right)^2}},
\end{equation}
where $S_i$ and $f_i$ represent observed and fitted sunspot 
numbers, respectively. Besides, we use the Anderson-Darling 
distance $A^2$ that is the basis of Anderson-Darling test to 
measure the goodness-of-fit, 
\begin{equation}
A^2=n\int_{-\infty}^{\infty} \frac{\left(F_n\left(t\right)-F\left(t\right)\right)^2}
  {F\left(t\right)\left(1-F\left(t\right)\right)} dF\left(t\right),
 \label{eq:Asquare}
\end{equation}
where $F$ is the theoretical distribution and $F_n$ is the
empirically observed distribution. Equation~(\ref{eq:Asquare}) shows 
that the Anderson-Darling distance $A^2$ places more weight on 
observations in the tails of the distribution. The procedure we 
calculate $A^2$ is as follows: Firstly, we should consider the curve 
of sunspot numbers as a distribution, i.e., the months $t_i$ from 
the beginning of a solar cycle is an observed sample value and $S_i$ 
or $f_i$ is the number of sample $t_i$. Secondly, we calculate the 
cumulative distribution function $\Phi$ according to the fitted curve. 
Lastly, we use the discrete formula to calculate the
Anderson-Darling distance
\begin{equation}
A^2=-n-\sum\limits_{j=1}^n\frac{2j-1}{n}\left[\ln{\left(\Phi\left(Y_j\right)\right)}+
  \ln{\left(1-\Phi\left(Y_{n+1-j}\right)\right)}\right].
 \label{eq:AsquareDiscrete}
\end{equation}
Here, $Y$ denotes sample value and $n$ is
the total number of samples.

The results of evaluating indices, $A^2$, $\sigma$, and 
$CC$, are also listed in Table~\ref{tab:4para}.
 
\subsection{Features of solar cycles}

For a solar cycle, the most important features are the cycle length 
$T_c$, the ascent time $T_a$, the descent time $T_d$, the solar 
maximum $S_m$, and the solar minimum $S_0$. In the following, we give 
the expressions of the features about the four logistic parameters.
Note that the units of $T_c$, $T_a$, and $T_d$ are years.
 
\begin{figure}
\epsscale{0.8} \plotone{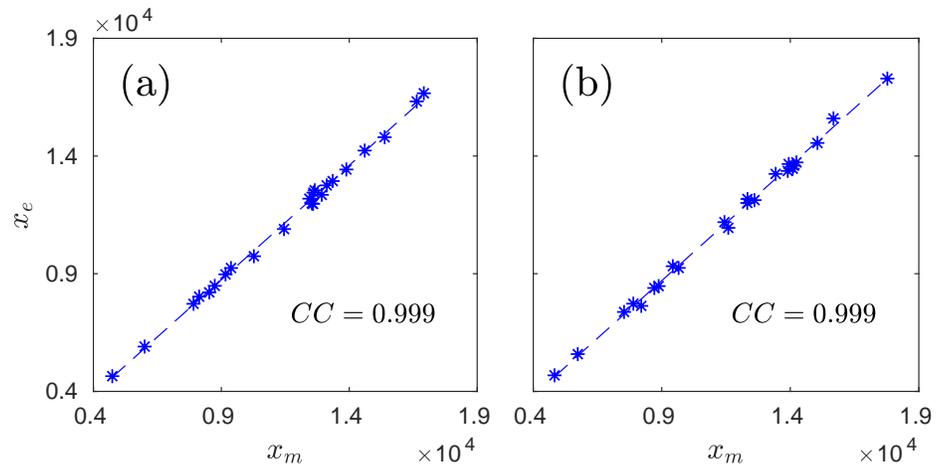}
\caption{$x_e$ as a function of $x_m$. Dashed line indicates the linear
fitting between $x_e$ and $x_m$ with correlation coefficient $CC=0.999$.
Panels (a) and (b) show the fitting results with the sunspot number
models of four- and two-parameter modified logistic differential 
equation, respectively.}
\label{fig:xeVsxm}
\end{figure}

Firstly, $S_m$ and $T_a$ can be derived by solving the equation $df/dt=0$,
\begin{eqnarray}
S_m&=&\frac{\alpha}{1+\alpha}\left(\frac{1}{1+\alpha}\right)^{1/\alpha}r_0x_m,\\
T_a&=& \left(12\alpha r_0\right)^{-1}\ln{\frac{x_m^\alpha/x_0^\alpha-1}{\alpha}}.
\end{eqnarray}
Secondly, when a cycle reaches its end, the variables $t$ and $x$ in 
Equation~(\ref{eq:x_int_mod}) becomes $t_e$ and $x_e$, respectively, 
where $x_e$ is the fitted cumulative sunspot number at the end of the 
cycle, and thus we can obtain the value of $t_e$ if $x_e$
is already known. Figure~\ref{fig:xeVsxm}(a) shows the relationship 
between the parameters $x_e$ and $x_m$ for solar cycles 1$-$23. 
The linear fitting is shown as a dashed line
\begin{equation}
x_e=0.9722x_m-6.93
\label{eq:xexm}
\end{equation}
with correlation coefficient $CC=0.999$ and significant at nearly 100\% 
confidence level. Therefore, $x_e$ can be estimated by the linear 
Equation~(\ref{eq:xexm}), and $T_c$ can be expressed as:
\begin{equation}
T_c\equiv \frac{t_e}{12}=\left(12\alpha r_0\right)^{-1}
 \ln{\frac{x_m^\alpha/x_0^\alpha-1}{x_m^\alpha/x_e^\alpha-1}}. \label{eq:T_c}
\end{equation}
Thirdly, $T_d$ can be obtained by subtracting $T_a$ from $T_c$,
\begin{equation}
T_d= \left(12\alpha r_0\right)^{-1}\ln{\frac{\alpha}{x_m^\alpha/x_e^\alpha-1}}.
\end{equation}
Lastly, combining the initial condition $x|_{t=0}=x_0$ with
Equation~(\ref{eq:x_diff_mod_re}), $S_0$ can be presented  as:
\begin{equation}
S_0=r_0\frac{x_0}{x_m}\frac{x_m^\alpha-x_0^\alpha}{x_m^{\alpha-1}}.
\label{eq:S_0}
\end{equation}

So that, if we have the fitting results of the four logistic parameters,
we can get the feature parameters of the solar cycles as shown above. The
absolute relative errors of $S_m$ and $T_a$ are listed in 
Table~\ref{tab:4para}, it can be seen that the relative error is generally 
in low level.
 
\subsection{Two-parameter function}

The variances and covariances denoted as 
$\sigma_{ij}$ $(i,j=\alpha, r_0, x_0, x_m)$ of the best fit
parameter values are also presented in Table~\ref{tab:4para}.
We can see that the covariance between parameters $\alpha$ and $r_0$ 
are large (negative) relative to their variances for all of the cycles.
Thus, an increase in $\alpha$ could lead to a decrease in $r_0$ without
changing the overall quality of the fit substantially, which makes
the parameters hard to interpret. Besides, the covariance between 
parameters $\alpha$ and $x_0$ are also large relative to their variances 
for some cycles. Therefore, the parameter $\alpha$ should be eliminated.

From Table~\ref{tab:4para} we can see that for the fitting results the 
mean and standard deviation of the asymmetry $\alpha$ are 0.317
and 0.339, respectively, and the mean and standard deviation 
of the maximum emergence rate $r_0$ are 0.479 and 0.480, 
respectively. Though for solar cycles $1-23$, $\alpha$ ranges from 
0.0274 to 1.28, we find that the fitting results are 
also very good if a suitable fixed value of $\alpha$ is chosen for all 
of the cycles. Next, we set $\alpha$ as a constant $0.2$ and a 
three-parameter function is obtained. After fitting the three-parameter 
function to the sunspot number data of solar cycles $1-23$, we get the 
mean and standard deviation of $r_0$ as 0.224 and 0.029, respectively, 
which indicates that the new model is more stable than the four-parameter 
one. Therefore, $r_0$ could be fixed to 0.224 for all of the cycles. 
Therefore, the four-parameter modified logistic function is reduced to a 
two-parameter one as
\begin{eqnarray}
f\left(x\right)&=&0.224 \frac{x}{x_m} \frac{x_m^{0.2}-x^{0.2}}{x_m^{-0.8}}
 \label{eq:x_diff_mod_2para}\\
x\left(t\right)&=&x_m {\left[1+\left(\frac{x_m^{0.2}}{x_0^{0.2}}-1\right)
 e^{-0.0448t}\right]^{-5}}. \label{eq:x_int_mod_2para}
\end{eqnarray}

We fit the two-parameter modified logistic differential equation to the 
sunspot number data of solar cycles $1-23$, the results are exhibited by 
the green curves in Figure~\ref{fig:solarCycles_fit}. It is shown that 
the two-parameter function can fit sunspot number data very well. 
Figure~\ref{fig:xeVsxm}(b) shows the relationship between the parameters 
$x_e$ and $x_m$ fitted by the two-parameter function for solar cycles
1$-$23. The linear fitting is shown as a dashed line
\begin{equation}
x_e=0.9778x_m-95.46
\label{eq:xexm2p}
\end{equation}
with correlation coefficient $CC=0.999$ and significant at nearly 100\% 
confidence level. Therefore, with two-parameter function,
$T_c$ can also be estimated by Equations~(\ref{eq:T_c}) and (\ref{eq:xexm2p}).
The fitting 
parameters and their uncertainties, the evaluating indices, and 
the absolute relative errors of 
$S_m$ and $T_a$ are listed in Table~\ref{tab:2para}. It can be seen that 
the fitting results of two-parameter function is similar to that of 
four-parameter one, so that the two-parameter function is also suitable 
for representing the sunspot number variations of solar cycles.
Table~\ref{tab:2para} also presented the variances and covariances 
of the two best fit parameter values, and we can find that the covariance
$\sigma_{x_0 x_m}$ is small for all of the cycles.
 
A model with less parameters is beneficial for prediction. On the one 
hand, when a solar cycle has progressed for 2 to 3 years, i.e., 
$\Delta t=2$ to 3 years, some models can be used to fit the data available 
to predict the behaviour of the remaining cycle \citep[e.g.,][]{Du2011, 
Hathaway2015}. In general, the model with fewer parameters would obtain 
better prediction results if there are less data available to fit. On the 
other hand, if we want to predict the sunspot number variations at the 
start of a solar cycle, only very few parameters can be estimated. 
Therefore, to reduce to two parameters is important for us to construct 
the prediction model based on the modified logistic differential equation.

\subsection{Comparison of fitting effects with other functions}

\citet{Schwarz1978} proposed the index $BIC$, the Bayesian 
information criterion or Schwarz criterion, with the lower value preferred,
to establish whether one model is significantly better than another,
\begin{equation}
BIC=\ln{\left(N\right)}k-2\ln{\hat{\left(L\right)}},
\label{eq:BIC}
\end{equation}
where $N$ is the number of months in a cycle while $k$ and $\hat{L}$ 
are the number of parameters and value of the likelihood function of
a model, respectively.

Table~\ref{tab:comparision} illustrates the fitting effects of
three kinds of modified logistic differential equations,
quasi-Plank function Equation~(\ref{eq:quasi_Planck}), and modified 
Gaussian function Equation~(\ref{eq:mod_Gaussian}) for fitting
Versions 1 and 2 sunspot number from cycles 1 to 23. Column 7 of 
Table~\ref{tab:comparision} shows the average of $BIC$.

Table \ref{tab:comparision} shows that the fitting effects of 
four-parameter modified logistic differential equation are similar with 
those of four-parameter modified Gaussian function, and the performance 
of three-parameter modified logistic differential equation is similar 
with that of three-parameter quasi-Plank function. In fact, the difference
between the fitting results of the five models is not significant.

\section{Prediction ability}
\label{sec:prediction}

\subsection{Prediction model}
\label{sec:prediction model}

In order to use the two-parameter function to predict the variations of 
sunspot numbers, we have to estimate the two parameters, i.e., $x_0$ 
and $x_m$.

\begin{figure}
\epsscale{0.5} \plotone{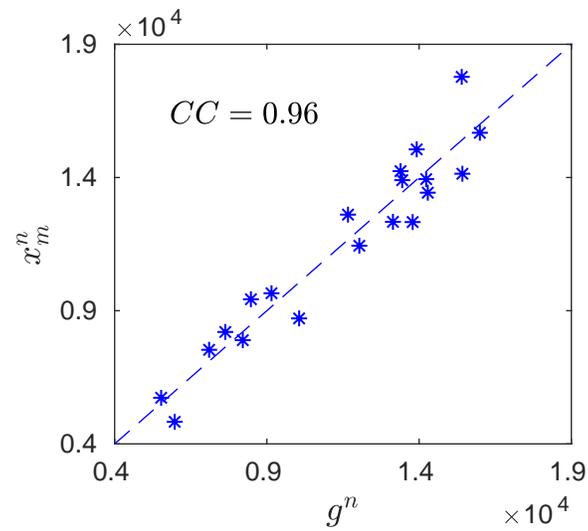}
\caption{Parameter $x_m$ as a function of $g^n$ which depends on the 
Shannon entropy values of the previous three cycles and the length 
of last cycle. Here, $n$ denotes cycle number.}
\label{fig:xm_prediction}
\end{figure}

For estimating $x_m$, we use Shannon entropy (also named as information 
entropy) as the potential predictor. Shannon entropy is proposed by 
Shannon \citep{Shannon1948} last century and applied to space physics 
recently \citep[e.g.,][]{LaurenzaEA2012, QinEA2013, KakadEA2015, 
KakadEA2017a}. It is a commonly used quantity to characterize the 
inherent randomness in the system. \citet{KakadEA2017a} divided every 
solar cycle to 5 phases each of which had an equal length of $T_c/5$, 
then they calculated the value of Shannon entropy for each phase using 
daily sunspot number. Based on the Shannon entropy, they predicted the 
maximum sunspot numbers for solar cycle 25. We follow 
\citealt{KakadEA2017a} to calculate the Shannon entropy. However, we 
use monthly sunspot number instead of daily one so that the 
calculation can be extended before solar cycle 10. The Shannon 
entropy values calculated by Version 2 sunspot number are listed 
in Table~\ref{tab:cycleFeatures} for cycles $1-23$, in which the Shannon 
entropy values are denoted as $E_i$ where $i=1-5$ denoting the 
phase number. The table also includes the data $T_c$ in the last column. 
Due to the fact that $S_m$ is highly correlated with $x_m$ ($r=0.97$ and 
significant at nearly 100\% confidence level for two-parameter fitting 
results) and the fact that $S_m$ is inversely proportional to the length 
of the previous cycle \citep{HathawayEA1994}, we use $E_i^{n-3}$, 
$E_i^{n-2}$, $E_i^{n-1}$, and $T_c^{n-1}$ as potential predictors to 
estimate $x_m^n$. Here the superscript denotes the cycle number. The 
stepwise regression method is used to deal with the multi-variable linear 
regression. We find that $x_m^n$ can be expressed as:
\begin{eqnarray}
x_m^n=g^n &\equiv& -3509 E_4^{n-3}+3097 E_2^{n-2}+4327 E_5^{n-2}
 -3190 E_1^{n-1}\nonumber\\
 &&+3189 E_4^{n-1}+1397 E_5^{n-1}-624 T_c^{n-1}-11862.
\label{eq:g}
\end{eqnarray}
Figure~\ref{fig:xm_prediction} shows the relationship between the fitted 
value of $x_m^n$ and $g^n$ from linear equation (\ref{eq:g}). The dashed 
line indicates the linear fitting between $x_m$ and $g^n$ with the 
correlation coefficient being $0.96$ and significant at nearly 100\% 
confidence level. Note that $g^n$ depends on the Shannon entropy in solar 
cycles $n-3$, $n-2$, and $n-1$, and $T_c$ in solar cycle $n-1$. Therefore, 
$x_m$ can be predicted at the start of the cycle.

The other parameter $x_0$ can be solved by taking the known quantities 
$S_0$, which is the sunspot number of the first month, i.e., solar 
minimum, and $x_m$ into Equation~(\ref{eq:S_0}). Note that, the 
value of $S_0$ of cycle 6 is 0, thus the corresponding $x_0$ is also
equals to 0, causing the sunspot number not to growth according
to Equation~(\ref{eq:x_diff_mod}). Therefore, $S_0$ of cycle 6 is 
assigned a value of 0.2, which is the minimum observed value except $0$ 
for all of the cycles. So that the sunspot number variations of a 
solar cycle can be predicted at the start of the cycle. The prediction 
model in this work is denoted as TMLP (Two-parameter Modified Logistic 
Prediction) model, hereafter.

\begin{figure}
\epsscale{1} \plotone{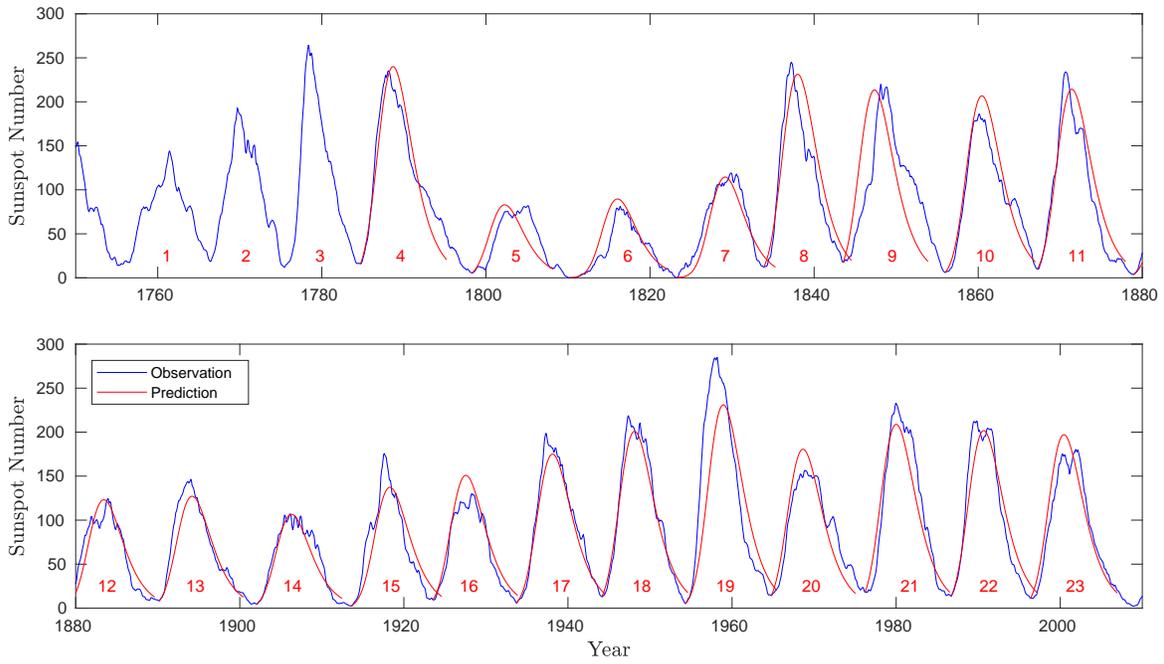}
\caption{Blue curves are the monthly smoothed sunspot number for cycles
 $1-23$, and the red curves are the TMLP predicting results based on the
 two-parameter modified logistic differential equation for cycles 
 $4-23$.}
\label{fig:cycles_pre}
\end{figure}

Figure~\ref{fig:cycles_pre} shows the prediction results by TMLP for 
Version 2 sunspot number from cycles $4-23$. The predicted 
cycle lengths are calculated by Equation~(\ref{eq:T_c}). The evaluating 
indices and the absolute relative error of $S_m$, $T_a$, and 
$T_c$ are exhibited in Table~\ref{tab:prediction}. It is noted that to
calculate the evaluating indices with Equations
(\ref{eq:sigma}$-$\ref{eq:Asquare}) the cycle length has to be correct, 
so the observed cycle length is used. From Figure~\ref{fig:cycles_pre} 
and Table~\ref{tab:prediction} we can see that the prediction results 
are good except for cycles 5, 9, and 19. The average relative error 
of $S_m$ is 8.8\% and the maximum one is 22\% in cycle 15.
For $T_a$, the average relative error is 17\% and the maximum one is 
71\% in cycle 5, during which the sunspot number increases in the 
beginning but decreases after about 1 year. If solar cycle 5 is removed 
the average relative error of $T_a$ is 14\%. The average relative 
error of $T_c$, which is not predicted by other fitting functions 
in the literature, is 9.5\%.
 
\subsection{Quasi-online determination of solar minimum moment}
\label{sec:quasi-online}
The above prediction is based on the condition that the solar 
minimum of a new cycle is already known, and thus the moment of solar 
minimum needs to be determined by using a quasi-online method. In fact, 
the goodness of prediction result strongly depends on the accurate 
determination of the moment and amplitude of the solar minimum. Therefore, 
we introduce a quasi-online method to determine the moment of solar 
minimum in the following.

\begin{figure}
\epsscale{0.9} \plotone{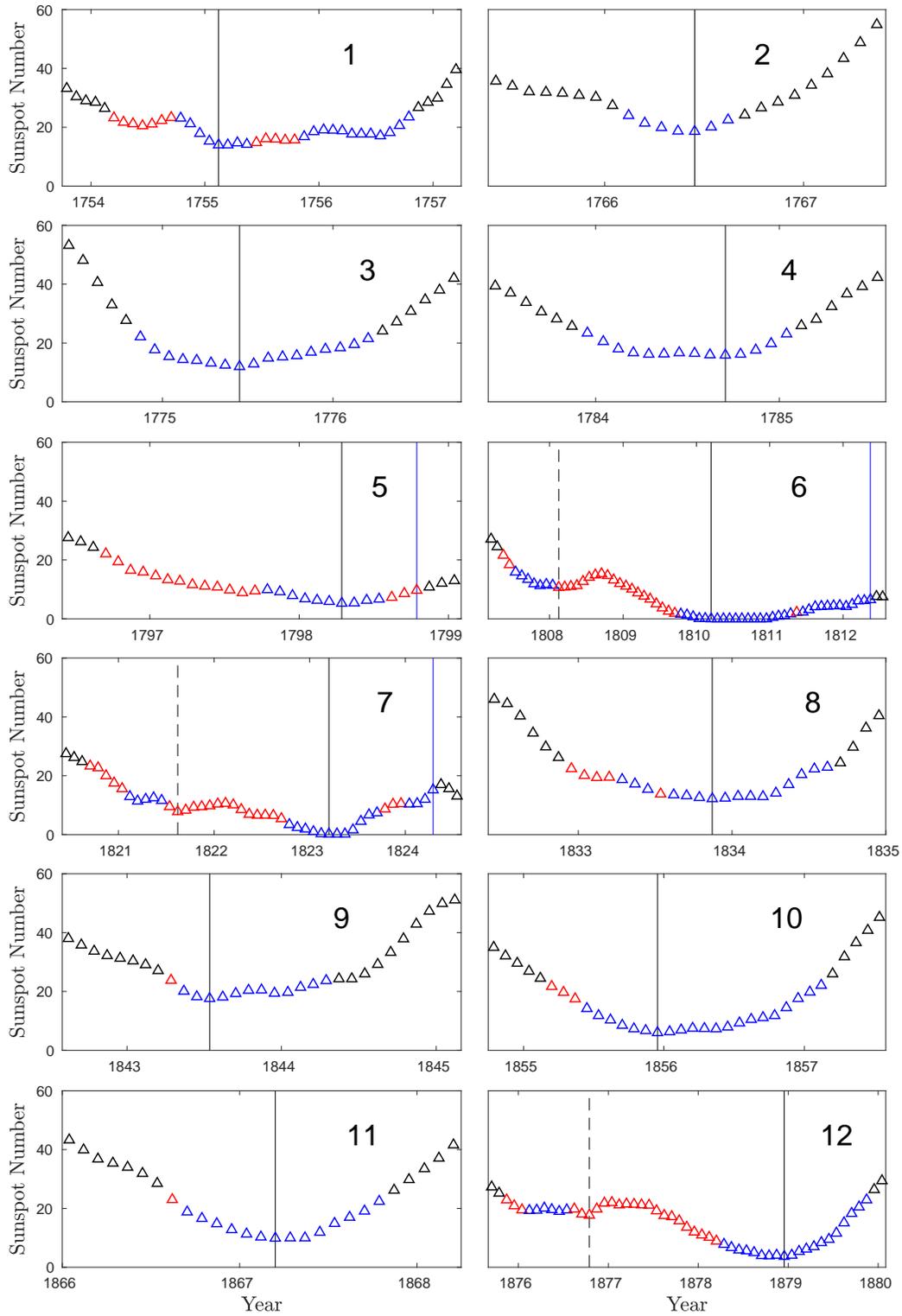}
\caption{Triangles are monthly smoothed sunspot numbers for cycles 
1$-$12, and the blue and red triangles are the potential periods of 
solar minimum. Red and blue indicate the trend of sunspot number is less 
and not less than 0. The black dashed and solid lines represent the 
potential solar minimum, and the blue lines indicate the 
position of the upper limit of $CI_{t_e}$.}
\label{fig:minima_1}
\end{figure}

\begin{figure}
\epsscale{0.9} \plotone{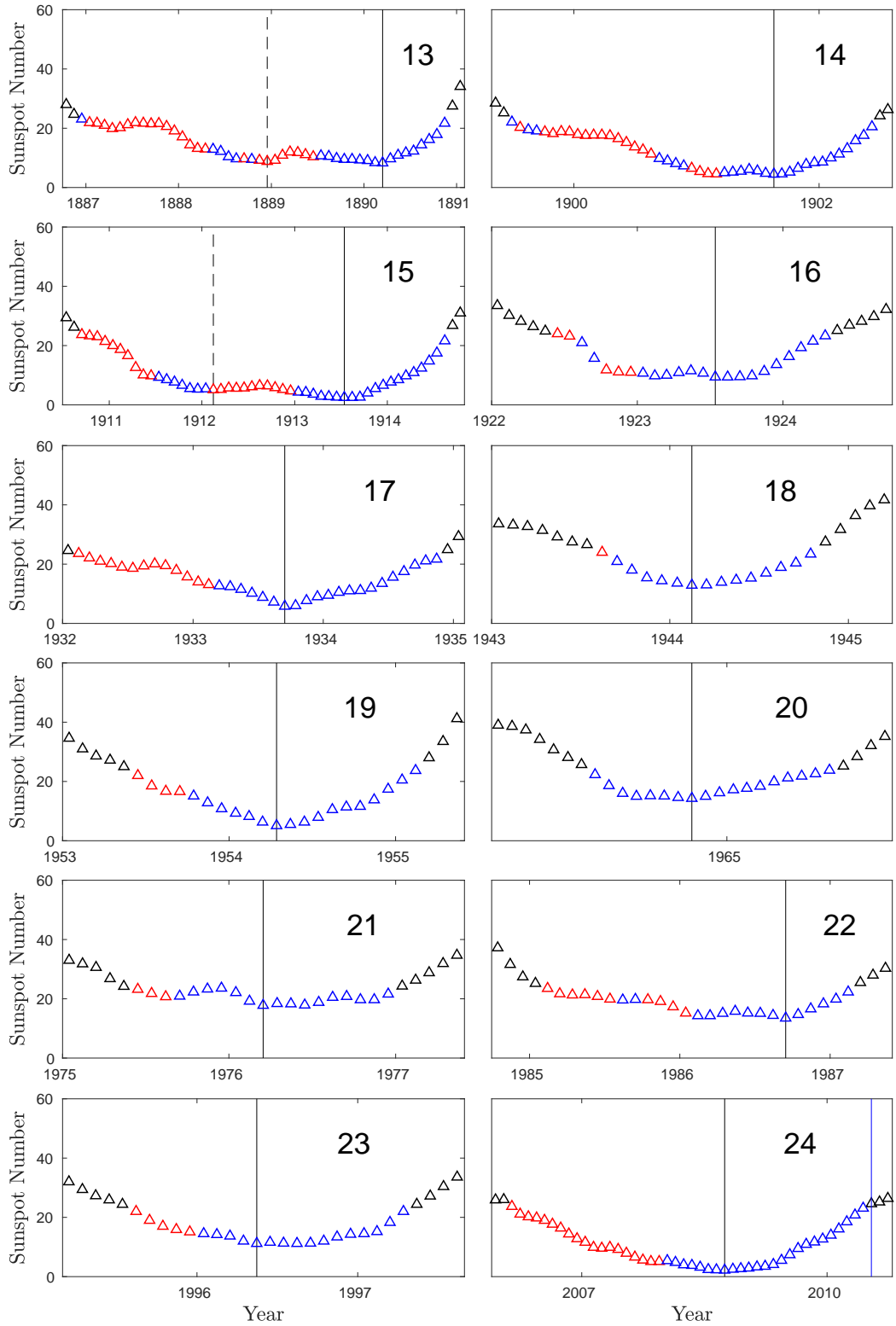}
\caption{Same as Figure~\ref{fig:minima_1} but for cycles 13$-$24.}
\label{fig:minima_2}
\end{figure}

Firstly, the mean and standard deviation of the amplitudes of solar 
minima for cycles 1$-$23 are 9.6 and 5.6 respectively, so that the confidence
interval (CI) of the amplitude of solar minimum at 99\% significant level 
is $CI_S=(-5, 24)$. Secondly, the mean and standard deviation of the 
end time
of cycles 1$-$23 are 132 months and 14 months, respectively, and 
thus the CI of end time at 99\% significant level is 
$CI_{t_e}=(95, 169)$. The blue and red triangles in Figures~\ref{fig:minima_1} 
and \ref{fig:minima_2} show the periods that meet the above two conditions 
for cycles 1$-$24, which can be called potential periods of solar minimum. 
The blue vertical lines in the figures indicate the position of the upper 
limit of $CI_{t_e}$.

We define that a potential solar minimum should have sunspot
number less than that in the previous months and less than or equal to
that in the next 8 months during the potential period of solar minimum. 
Based on the standards, the potential minima are obtained and presented 
by the black dashed and solid lines in Figures~\ref{fig:minima_1} and 
\ref{fig:minima_2}, where the solid lines indicate the solar
minima while the dashed lines indicate the local minima but not the
solar minima.

In order to determine whether a potential minimum is a solar 
minimum, we use the trend of sunspot number, which is defined as the 
slope of the linear fitting for the current month and the next 14 months 
monthly sunspot number. In Figures~\ref{fig:minima_1} and \ref{fig:minima_2}, 
the blue triangles indicate that the trend is greater than or equal to 0, and
the red triangles denote that the trend is less than 0. We can find that the 
trends of all of the black dashed lines are less than 0 while those of black 
solid lines are greater than or equal to 0. Therefore, a potential minimum is
determined as a solar minimum if the trend is greater than or equal to 0. This 
is a quasi-online method from which we can obtain the moment of solar minimum 
of a new cycle 14 months after the solar minimum. For Version 1 sunspot number, 
the method is valid for solar cycles 1-24.

\subsection{Comparison of prediction ability with other functions}

Since the prediction models established by \citet{HathawayEA1994}
and \citet{Du2011} are based on Version 1 sunspot number, we adopt the
above procedure in subsections \ref{sec:prediction model} and 
\ref{sec:quasi-online} to establish a new prediction model with
Version 1 sunspot number for comparison purpose. The equations corresponding 
to Equations~(\ref{eq:xexm2p}) and (\ref{eq:g}) are given by:
\begin{equation}
x_e=0.9750x_m-40.13
\label{eq:xexm2p_v1}
\end{equation}
and
\begin{eqnarray}
x_m^n=g^n &\equiv& 1793 E_2^{n-2}-3198 E_4^{n-2}+2457 E_5^{n-2}
 \nonumber\\
 &&+2472 E_4^{n-1}+2302 E_5^{n-1}-747 T_c^{n-1}-12297.
\label{eq:g_v1}
\end{eqnarray}
The prediction results are exhibited in Table~\ref{tab:prediction_v1}.
Note that, \citet{HathawayEA1994} predicted the amplitude $a$ rather than
the sunspot maximum, and the predicted $a$ and starting time of a solar
cycle are used to predict the rest behavior of the cycle. The comparison
shows that our prediction model can obtain a good prediction result earlier 
than other models for predicting Version 1 sunspot number. What's more, our
prediction model can predict the cycle length simultaneously.

\subsection{Prediction of solar cycle 24}

Solar cycle 24 starts in December 2008 and has progressed $\sim9$ years,
which is long enough so that the parameters of the fitting function can 
be determined with high accuracy 
\citep[e.g.,][]{HathawayEA1994, Du2011}. Thus, 
Equations~(\ref{eq:quasi_Planck}), (\ref{eq:mod_Gaussian}) and 
(\ref{eq:x_diff_mod_re}) can be used to estimate the sunspot number 
in the remaining part of the cycle by fitting the $\sim9$ years data 
of cycle 24, and the estimated value can be used to compare
with the prediction result by TMLP.

\begin{figure}
\epsscale{0.7} \plotone{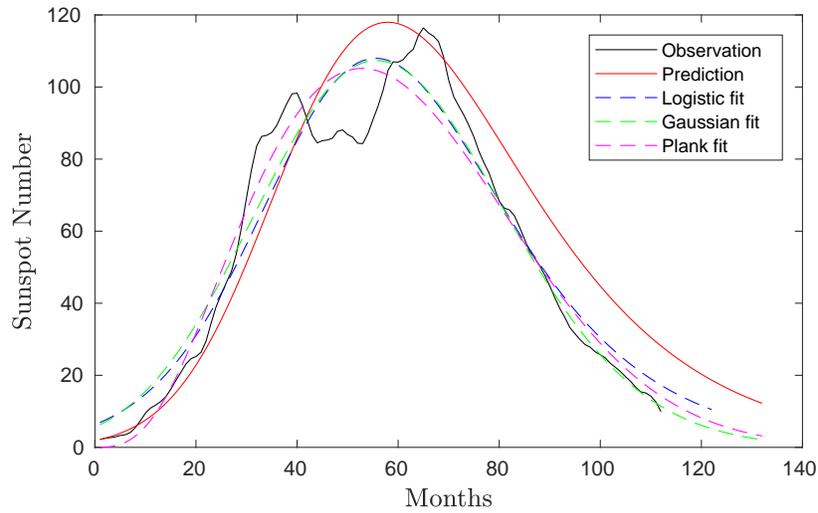}
\caption{Monthly smoothed sunspot number curve (black line), the fitting
 results by the four-parameter modified logistic differential equation 
 (blue dashed line), the four-parameter modified Gaussian function 
 (green dashed line), the three-parameter quasi-Plank function 
 (pink dashed line), and the TMLP predicting result (red line), for 
 cycle 24. The cycle lengths of the red line and the blue dashed line 
 are obtained by Equation~(\ref{eq:T_c}) while those of the 
 green and pink dashed  line are assigned to 11 years.}
\label{fig:cycle24_pre}
\end{figure}

The red and black curves in Figure~\ref{fig:cycle24_pre} show the TMLP
result and the observation for solar cycle 24, respectively. The 
pink, green and blue dashed curves indicate the fitting results 
by the quasi-Plank function Equation~(\ref{eq:quasi_Planck})
\citep{HathawayEA1994}, modified Gaussian function 
Equation~(\ref{eq:mod_Gaussian}) \citep{Du2011} and the four-parameter 
modified logistic differential equation Equation~(\ref{eq:x_diff_mod_re}), 
respectively. The cycle length from Equation~(\ref{eq:T_c}) by fitting 
Equation~(\ref{eq:x_diff_mod_re}) to $\sim9$ years observed data and 
that of the prediction result by TMLP are 10.1 and 
11.0 years, respectively. Note that, the cycle lengths of 
Equation~(\ref{eq:quasi_Planck}) and 
Equation~(\ref{eq:mod_Gaussian}) are taken as 11 years. 
In order to estimate the error of prediction result, the 
CI of prediction result at 95\% significant level should be obtained 
by calculating the prediction results of cycles 4$-$23. The mean and 
standard error of the prediction results of cycles 4$-$23 are 
-3.3\% and 10.7\%, respectively, thus the CI at 95\% significant 
level are (-24.3\%, 17.7\%). Therefore, the predicted cycle length 
of cycle 24 by the TMLP is 11.0 (95\% CI is 8.3 to 12.9) years.

The TMLP prediction result is consistent with the observation for the 
first 40 months of the cycle, and the sunspot number from TMLP prediction 
is higher than other curves after about 70 months but without larger error, 
i.e., the relative errors of $S_m$ and $T_a$ are 1.4\% and 
12\%, respectively.

\section{Conclusions}
\label{sec:conclusion}

In this paper, we use the logistic function to study the time profile of 
monthly smoothed sunspot number. Due to the fact that the 
variations of sunspot number is asymmetrical in cycles, we introduce an 
asymmetry $\alpha$ to modify the logistic function. There are three 
parameters in addition in the function, namely, maximum cumulative 
sunspot number or total sunspot number $x_m$, initial cumulative sunspot 
number $x_0$, and maximum emergence rate $r_0$. By using these parameters 
we can give the features of solar cycles. 
We find that the modified logistic differential equation can fit the 
variations of sunspot number of solar cycles very well. 
The fitting results show 
that if we choose 
$\alpha=0.2$ and $r_0=0.224$, the four-parameter function is reduced to 
a two-parameter one.

Although the fitting results of the four-parameter function are 
slightly better than those of the two-parameter one, it is hard to 
establish a prediction model of sunspot number with four parameters. 
Therefore, we use the two-parameter modified logistic differential 
equation to construct the prediction model of the sunspot number in 
solar cycles.
At the start of a cycle the parameter $x_m$ is estimated with a linear 
expression of the Shannon entropy values of the three previous 
cycles and the last cycle length. The correlation coefficient of the 
linear regression is 0.96, and the regression is significant at nearly 
100\% confidence level. The other parameter $x_0$ could be solved by 
taking the known quantities $S_0$, the sunspot number of the first month, 
and $x_m$ into Equation~(\ref{eq:S_0}). Therefore, the sunspot number 
variations in a solar cycle can be predicted at the start of the cycle by 
the two-parameter modified logistic differential equation if the
solar minimum is already known. Furthermore, we introduce a quasi-online
method to determine the moment of solar minimum when 14 months after
a new cycle starts. The prediction model is called TMLP in this paper, 
and its predicting results of solar cycles $4-23$ are all good 
except for cycles 5, 9, and 19. For Version 2 sunspot number, the
average relative error of the solar maximum is 8.8\%. The average 
relative errors of the ascent time is 17\%, however, it would be reduced 
to 14\% if cycle 5 is removed. In addition, the average relative error of 
cycle length, which is not predicted by other fitting functions in the 
literature, is 9.5\%. In order to compare the prediction
ability of TMLP with other fitting functions, we also establish a prediction
model and obtain its prediction results for Version 1 sunspot number.
The comparison shows that TMLP can obtain a good prediction result
earlier than other fitting functions and predict the cycle length
simultaneously.
 
Furthermore, using the TMLP, we get the prediction of the sunspot number 
variations of solar cycle 24, which is compared with the observation and 
the fitting results by the three-parameter quasi-Plank function
\citep{HathawayEA1994}, four-parameter modified Gaussian function 
\citep{Du2011} and the four-parameter modified logistic differential 
equation. The comparison indicates that the sunspot number predicted
by TMLP is consistent with the observation at the first 40 months, but is 
relatively larger than other curves after about 70 months. The relative 
errors of the solar maximum and ascent time are 1.4\% and 
12\%, respectively. The TMLP model also predicts that the 
length of cycle 24 is 11.0 (95\% CI is 8.3 to 12.9)
years, which is slightly larger than the fitting results of the 
four-parameter modified logistic differential equation. 
The predicted cycle length of cycle 24 is similar with other
works, such as 11.33 years of \citet{UzalEA2012}, 11.3 years of
\citet{Pishkalo2014}, and 11.01 years of \citet{KakadEA2017b}.

We plan to predict sunspot number variations of solar cycle 25 
when we can determine the solar minimum, i.e., 14 months after 
the solar minimum of the cycle.

\acknowledgments
This work was supported, in part, under grants NNSFC 41874206 and 
NNSFC 41574172. We thank the SIDC and SILSO teams and the Royal 
Observatory of Belgium for international sunspot data.


\listofchanges

\clearpage

\clearpage

\clearpage

\clearpage
 
\clearpage
  
\clearpage

\clearpage
  
\clearpage

\clearpage

\clearpage
\begin{longrotatetable}
\setlength{\tabcolsep}{1mm}{
\begin{deluxetable}{ccccccccccccccccccccc}
\tabletypesize{\tiny}
\tablecaption{Fitted parameters of the four-parameter modified
logistic differential equation, the variances and covariances
of these parameters, the evaluating indices, and the absolute
relative error of cycle features for cycles $1-23$.
\label{tab:4para}}
\tablehead{
\colhead{Cycle No.} & \colhead{$\alpha$} & \colhead{$r_0$} & \colhead{$x_0$} &
\colhead{$x_m$} & 
\colhead{$\sigma^2_{\alpha\alpha}$} & \colhead{$\sigma^2_{r_0 r_0}$} & \colhead{$\sigma^2_{x_0 x_0}$} & \colhead{$\sigma^2_{x_m x_m}$} &
\colhead{$\sigma^2_{\alpha r_0}$} & \colhead{$\sigma^2_{\alpha x_0}$} & \colhead{$\sigma^2_{\alpha x_m}$} &
\colhead{$\sigma^2_{r_0 x_0}$} & \colhead{$\sigma^2_{r_0 x_m}$} & \colhead{$\sigma^2_{x_0 x_m}$} &
\colhead{$x_e$} & \colhead{$A^2$} & \colhead{$\sigma$} &
\colhead{$CC$} & \colhead{$\frac{\delta S_m}{S_m}$} & \colhead{$\frac{\delta T_a}{T_a}$}
}
\startdata
1 & 9.12 $\pm$ 1.17 E-1 & 5.06 $\pm$ 0.46 E-2 & 2.74 $\pm$ 0.48 E+2 & 1.03 $\pm$ 0.01 E+4 & 1.4 E-2 & 2.1 E-5 & 2.3 E+3 & 1.7 E+4 & -5.3 E-4 & 5.2 E+0 & 3.5 E-1 & -2.1 E-1 & -8.6 E-2 & 2.0 E+3 & 9.74 E+3 & 71 & 8.5 & 0.973 & 15\% & 4.1\% \\
2 & 2.64 $\pm$ 0.61 E-1 & 1.79 $\pm$ 0.37 E-1 & 1.49 $\pm$ 0.34 E+2 & 1.15 $\pm$ 0.01 E+4 & 3.8 E-3 & 1.3 E-3 & 1.2 E+3 & 1.7 E+4 & -2.2 E-3 & 1.9 E+0 & -2.1 E-1 & -1.2 E+0 & -8.7 E-2 & 9.8 E+2 & 1.09 E+4 & 66 & 9.8 & 0.983 & 8.9\% & 11\% \\
3 & 3.49 $\pm$ 0.64 E-2 & 1.57 $\pm$ 0.28 E+0 & 5.65 $\pm$ 1.35 E+0 & 1.24 $\pm$ 0.01 E+4 & 4.0 E-5 & 7.8 E-2 & 1.8 E+0 & 2.1 E+4 & -1.8 E-3 & 3.9 E-3 & -1.7 E-2 & -2.0 E-1 & -1.3 E+0 & 1.0 E+2 & 1.22 E+4 & 16 & 13 & 0.987 & 7.1\% & 12\% \\
4 & 2.74 $\pm$ 0.41 E-2 & 1.31 $\pm$ 0.19 E+0 & 1.34 $\pm$ 0.15 E+2 & 1.69 $\pm$ 0.01 E+4 & 1.7 E-5 & 3.7 E-2 & 2.2 E+2 & 3.7 E+4 & -7.9 E-4 & 1.3 E-2 & -1.5 E-2 & -8.3 E-1 & -1.6 E+0 & 1.7 E+3 & 1.67 E+4 & 58 & 14 & 0.982 & 6.3\% & 11\% \\
5 & 8.78 $\pm$ 1.38 E-1 & 6.11 $\pm$ 0.70 E-2 & 1.16 $\pm$ 0.29 E+2 & 6.00 $\pm$ 0.08 E+3 & 1.9 E-2 & 4.9 E-5 & 8.2 E+2 & 7.7 E+3 & -9.5 E-4 & 3.7 E+0 & 1.9 E+0 & -2.0 E-1 & -1.5 E-1 & 9.3 E+2 & 5.90 E+3 & 25 & 6.6 & 0.973 & 2.0\% & 24\% \\
6 & 4.30 $\pm$ 0.66 E-1 & 1.19 $\pm$ 0.15 E-1 & 2.40 $\pm$ 1.39 E+0 & 4.74 $\pm$ 0.06 E+3 & 4.4 E-3 & 2.4 E-4 & 1.9 E+0 & 4.3 E+3 & -1.0 E-3 & 8.7 E-2 & 2.5 E-1 & -2.1 E-2 & -1.4 E-1 & 2.3 E+1 & 4.65 E+3 & 8.9 & 5.1 & 0.979 & 8.9\% & 6.3\% \\
7 & 1.28 $\pm$ 0.14 E+0 & 4.84 $\pm$ 0.33 E-2 & 1.79 $\pm$ 0.31 E+2 & 8.53 $\pm$ 0.10 E+3 & 1.8 E-2 & 1.1 E-5 & 9.3 E+2 & 1.0 E+4 & -4.4 E-4 & 3.8 E+0 & 9.9 E-1 & -9.9 E-2 & -6.8 E-2 & 1.0 E+3 & 8.21 E+3 & 54 & 7.4 & 0.982 & 2.0\% & 5.2\% \\
8 & 4.07 $\pm$ 0.84 E-2 & 1.15 $\pm$ 0.23 E+0 & 2.30 $\pm$ 0.41 E+1 & 1.34 $\pm$ 0.01 E+4 & 7.1 E-5 & 5.4 E-2 & 1.7 E+1 & 2.2 E+4 & -2.0 E-3 & 1.7 E-2 & -4.6 E-2 & -5.3 E-1 & -2.2 E-1 & 3.1 E+2 & 1.29 E+4 & 25 & 12 & 0.985 & 7.9\% & 4.7\% \\
9 & 2.24 $\pm$ 0.58 E-1 & 1.68 $\pm$ 0.39 E-1 & 6.49 $\pm$ 2.40 E+1 & 1.54 $\pm$ 0.02 E+4 & 3.3 E-3 & 1.5 E-3 & 5.8 E+2 & 4.5 E+4 & -2.2 E-3 & 1.3 E+0 & -5.1 E-1 & -8.9 E-1 & -6.5 E-2 & 1.1 E+3 & 1.48 E+4 & 63 & 15 & 0.971 & 13\% & 13\% \\
10 & 3.14 $\pm$ 0.63 E-2 & 1.20 $\pm$ 0.24 E+0 & 1.57 $\pm$ 0.30 E+1 & 1.29 $\pm$ 0.01 E+4 & 4.0 E-5 & 5.6 E-2 & 9.0 E+0 & 2.3 E+4 & -1.5 E-3 & 7.2 E-3 & -4.7 E-2 & -3.1 E-1 & 3.5 E-2 & 2.5 E+2 & 1.24 E+4 & 44 & 11 & 0.980 & 5.0\% & 5.6\% \\
11 & 1.60 $\pm$ 0.39 E-1 & 3.12 $\pm$ 0.71 E-1 & 3.82 $\pm$ 1.17 E+1 & 1.27 $\pm$ 0.01 E+4 & 1.5 E-3 & 5.0 E-3 & 1.4 E+2 & 1.5 E+4 & -2.8 E-3 & 4.3 E-1 & -1.1 E-1 & -7.9 E-1 & -8.9 E-2 & 2.9 E+2 & 1.25 E+4 & 6.7 & 10 & 0.990 & 8.2\% & 8.7\% \\
12 & 7.19 $\pm$ 1.56 E-1 & 7.29 $\pm$ 1.21 E-2 & 2.43 $\pm$ 0.63 E+2 & 8.13 $\pm$ 0.13 E+3 & 2.4 E-2 & 1.4 E-4 & 3.9 E+3 & 1.7 E+4 & -1.9 E-3 & 9.2 E+0 & 3.7 E+0 & -7.3 E-1 & -3.9 E-1 & 3.2 E+3 & 8.03 E+3 & 43 & 9.7 & 0.970 & 6.1\% & 13\% \\
13 & 4.08 $\pm$ 0.46 E-2 & 1.03 $\pm$ 0.11 E+0 & 5.28 $\pm$ 0.37 E+1 & 9.38 $\pm$ 0.05 E+3 & 2.1 E-5 & 1.3 E-2 & 1.4 E+1 & 3.2 E+3 & -5.1 E-4 & 7.0 E-3 & -6.4 E-3 & -2.0 E-1 & -1.0 E-1 & 1.1 E+2 & 9.24 E+3 & 11 & 4.4 & 0.995 & 3.0\% & 9.3\% \\
14 & 3.66 $\pm$ 0.97 E-1 & 1.20 $\pm$ 0.27 E-1 & 1.06 $\pm$ 0.34 E+2 & 7.91 $\pm$ 0.11 E+3 & 9.1 E-3 & 7.2 E-4 & 1.1 E+3 & 1.2 E+4 & -2.6 E-3 & 3.0 E+0 & 1.0 E+0 & -8.7 E-1 & -4.2 E-1 & 1.1 E+3 & 7.72 E+3 & 49 & 8.6 & 0.973 & 1.3\% & 9.1\% \\
15 & 4.86 $\pm$ 0.80 E-1 & 1.16 $\pm$ 0.15 E-1 & 8.27 $\pm$ 2.22 E+1 & 9.16 $\pm$ 0.12 E+3 & 6.3 E-3 & 2.3 E-4 & 4.9 E+2 & 1.6 E+4 & -1.2 E-3 & 1.6 E+0 & -4.2 E-1 & -3.3 E-1 & -7.2 E-2 & 5.9 E+2 & 8.97 E+3 & 18 & 10 & 0.980 & 13\% & 5.7\% \\
16 & 4.02 $\pm$ 0.85 E-1 & 1.21 $\pm$ 0.22 E-1 & 1.22 $\pm$ 0.32 E+2 & 8.74 $\pm$ 0.10 E+3 & 7.2 E-3 & 4.6 E-4 & 1.0 E+3 & 1.1 E+4 & -1.8 E-3 & 2.6 E+0 & 4.4 E-1 & -6.6 E-1 & -2.2 E-1 & 8.2 E+2 & 8.48 E+3 & 38 & 8.1 & 0.982 & 0.39\% & 14\% \\
17 & 5.24 $\pm$ 0.77 E-2 & 8.06 $\pm$ 1.15 E-1 & 9.29 $\pm$ 1.41 E+0 & 1.25 $\pm$ 0.00 E+4 & 5.9 E-5 & 1.3 E-2 & 2.0 E+0 & 7.3 E+3 & -8.9 E-4 & 7.5 E-3 & -3.8 E-2 & -1.2 E-1 & 2.0 E-1 & 4.8 E+1 & 1.20 E+4 & 57 & 6.6 & 0.994 & 4.4\% & 15\% \\
18 & 1.25 $\pm$ 0.34 E-1 & 3.63 $\pm$ 0.91 E-1 & 4.00 $\pm$ 1.12 E+1 & 1.39 $\pm$ 0.01 E+4 & 1.1 E-3 & 8.3 E-3 & 1.2 E+2 & 1.5 E+4 & -3.0 E-3 & 3.5 E-1 & -2.2 E-1 & -9.6 E-1 & 3.1 E-1 & 2.2 E+2 & 1.34 E+4 & 58 & 9.7 & 0.991 & 0.020\% & 17\% \\
19 & 6.94 $\pm$ 0.86 E-2 & 6.97 $\pm$ 0.83 E-1 & 1.95 $\pm$ 0.24 E+1 & 1.66 $\pm$ 0.00 E+4 & 7.4 E-5 & 6.9 E-3 & 5.6 E+0 & 7.3 E+3 & -7.1 E-4 & 1.6 E-2 & -4.1 E-2 & -1.6 E-1 & 1.8 E-1 & 6.2 E+1 & 1.63 E+4 & 18 & 7.3 & 0.997 & 0.45\% & 6.7\% \\
20 & 4.44 $\pm$ 0.82 E-2 & 7.87 $\pm$ 1.43 E-1 & 5.91 $\pm$ 0.63 E+1 & 1.26 $\pm$ 0.01 E+4 & 6.8 E-5 & 2.0 E-2 & 4.0 E+1 & 1.1 E+4 & -1.2 E-3 & 2.7 E-2 & -4.3 E-2 & -5.1 E-1 & 2.2 E-1 & 3.2 E+2 & 1.20 E+4 & 65 & 7.3 & 0.989 & 1.0\% & 5.7\% \\
21 & 2.29 $\pm$ 0.43 E-1 & 2.11 $\pm$ 0.35 E-1 & 5.31 $\pm$ 1.48 E+1 & 1.46 $\pm$ 0.01 E+4 & 1.8 E-3 & 1.2 E-3 & 2.2 E+2 & 1.5 E+4 & -1.5 E-3 & 5.9 E-1 & -4.5 E-2 & -5.0 E-1 & -1.2 E-1 & 3.2 E+2 & 1.42 E+4 & 48 & 10 & 0.992 & 0.22\% & 9.8\% \\
22 & 1.20 $\pm$ 0.41 E-1 & 4.06 $\pm$ 1.30 E-1 & 4.88 $\pm$ 1.51 E+1 & 1.31 $\pm$ 0.01 E+4 & 1.6 E-3 & 1.7 E-2 & 2.3 E+2 & 1.9 E+4 & -5.3 E-3 & 5.7 E-1 & -2.5 E-1 & -1.8 E+0 & 3.7 E-1 & 3.8 E+2 & 1.28 E+4 & 42 & 12 & 0.987 & 4.1\% & 11\% \\
23 & 3.63 $\pm$ 0.47 E-1 & 1.27 $\pm$ 0.14 E-1 & 1.03 $\pm$ 0.20 E+2 & 1.26 $\pm$ 0.01 E+4 & 2.2 E-3 & 2.0 E-4 & 3.9 E+2 & 1.0 E+4 & -6.6 E-4 & 8.7 E-1 & 7.5 E-2 & -2.7 E-1 & -9.7 E-2 & 4.5 E+2 & 1.24 E+4 & 20 & 7.9 & 0.992 & 0.70\% & 18\% \\
\hline
avg. & 3.17 E-1 & 4.79 E-1 &  &  &  &  &  &  &  &  &  &  &  &  &  & 39 & 9.3 & 0.984 & 5.2\% & 10\% \\
s.d. & 3.39 E-1 & 4.80 E-1 &  &  &  &  &  &  &  &  &  &  &  &  &  & 21 & 2.6 & 0.008 & 4.5\% & 4.9\%
\enddata
\end{deluxetable}
}
\end{longrotatetable}

\clearpage
\begin{deluxetable}{cccccccccccc}
\tabletypesize{\footnotesize}
\tablecaption{Same as Table~\ref{tab:4para} but for the two-parameter
 function.
\label{tab:2para}}
\tablehead{
\colhead{Cycle No.} & \colhead{$x_0$} & \colhead{$x_m$}  & 
\colhead{$\sigma_{x_0 x_0}$} & \colhead{$\sigma_{x_m x_m}$} & \colhead{$\sigma_{x_0 x_m}$} &
\colhead{$x_e$} &
\colhead{$A^2$} & \colhead{$\sigma$} & \colhead{$CC$} &
\colhead{$\frac{\delta S_m}{S_m}$} & \colhead{$\frac{\delta T_a}{T_a}$}
}
\startdata
1 & 2.50 $\pm$ 0.37 E+0 & 8.88 $\pm$ 0.16 E+3 & 1.2 E-1 & 2.3 E+4 & 6.1 E+0 & 8.47 E+3 & 191 & 17 & 0.890 & 7.6\% & 12\% \\
2 & 1.33 $\pm$ 0.06 E+2 & 1.16 $\pm$ 0.01 E+4 & 3.1 E+1 & 1.0 E+4 & 1.1 E+2 & 1.09 E+4 & 93 & 10 & 0.982 & 9.9\% & 11\% \\
3 & 2.05 $\pm$ 0.17 E+2 & 1.41 $\pm$ 0.02 E+4 & 2.6 E+2 & 6.8 E+4 & 9.8 E+2 & 1.35 E+4 & 219 & 27 & 0.941 & 20\% & 16\% \\
4 & 1.40 $\pm$ 0.09 E+2 & 1.57 $\pm$ 0.01 E+4 & 6.8 E+1 & 3.2 E+4 & 2.9 E+2 & 1.56 E+4 & 347 & 20 & 0.961 & 0.027\% & 11\% \\
5 & 4.64 $\pm$ 0.43 E+0 & 5.73 $\pm$ 0.07 E+3 & 1.8 E-1 & 6.2 E+3 & 4.9 E+0 & 5.58 E+3 & 46 & 7.9 & 0.961 & 4.8\% & 32\% \\
6 & 1.86 $\pm$ 0.15 E-1 & 4.83 $\pm$ 0.05 E+3 & 2.2 E-4 & 2.9 E+3 & 1.1 E-1 & 4.68 E+3 & 36 & 5.4 & 0.977 & 11\% & 5.1\% \\
7 & 1.66 $\pm$ 0.17 E+0 & 8.20 $\pm$ 0.11 E+3 & 2.9 E-2 & 1.3 E+4 & 2.3 E+0 & 7.63 E+3 & 185 & 12 & 0.957 & 3.2\% & 16\% \\
8 & 1.62 $\pm$ 0.09 E+2 & 1.41 $\pm$ 0.01 E+4 & 8.2 E+1 & 3.0 E+4 & 3.4 E+2 & 1.36 E+4 & 98 & 18 & 0.970 & 13\% & 8.9\% \\
9 & 1.06 $\pm$ 0.10 E+1 & 1.39 $\pm$ 0.01 E+4 & 8.3 E-1 & 3.2 E+4 & 2.4 E+1 & 1.37 E+4 & 84 & 19 & 0.949 & 4.9\% & 11\% \\
10 & 3.48 $\pm$ 0.23 E+1 & 1.23 $\pm$ 0.01 E+4 & 5.1 E+0 & 1.9 E+4 & 5.4 E+1 & 1.20 E+4 & 46 & 14 & 0.967 & 0.68\% & 7.3\% \\
11 & 1.18 $\pm$ 0.06 E+2 & 1.34 $\pm$ 0.01 E+4 & 3.1 E+1 & 1.7 E+4 & 1.5 E+2 & 1.32 E+4 & 63 & 14 & 0.982 & 14\% & 11\% \\
12 & 4.55 $\pm$ 0.34 E+1 & 7.89 $\pm$ 0.10 E+3 & 1.2 E+1 & 1.1 E+4 & 6.4 E+1 & 7.73 E+3 & 27 & 11 & 0.964 & 4.8\% & 22\% \\
13 & 1.26 $\pm$ 0.04 E+2 & 9.43 $\pm$ 0.05 E+3 & 1.4 E+1 & 3.4 E+3 & 4.7 E+1 & 9.32 E+3 & 30 & 5.9 & 0.992 & 3.4\% & 6.8\% \\
14 & 2.88 $\pm$ 0.21 E+1 & 7.53 $\pm$ 0.09 E+3 & 4.6 E+0 & 9.1 E+3 & 3.5 E+1 & 7.38 E+3 & 30 & 9.6 & 0.966 & 5.5\% & 5.7\% \\
15 & 4.80 $\pm$ 0.31 E+1 & 9.66 $\pm$ 0.12 E+3 & 9.3 E+0 & 1.4 E+4 & 6.7 E+1 & 9.24 E+3 & 89 & 12 & 0.971 & 18\% & 2.0\% \\
16 & 5.44 $\pm$ 0.28 E+1 & 8.71 $\pm$ 0.08 E+3 & 7.6 E+0 & 6.9 E+3 & 4.2 E+1 & 8.39 E+3 & 49 & 8.3 & 0.981 & 0.36\% & 16\% \\
17 & 4.36 $\pm$ 0.15 E+1 & 1.26 $\pm$ 0.00 E+4 & 2.2 E+0 & 6.3 E+3 & 2.1 E+1 & 1.21 E+4 & 45 & 7.9 & 0.992 & 4.8\% & 19\% \\
18 & 9.36 $\pm$ 0.36 E+1 & 1.42 $\pm$ 0.01 E+4 & 1.2 E+1 & 1.1 E+4 & 7.0 E+1 & 1.37 E+4 & 86 & 11 & 0.989 & 2.3\% & 18\% \\
19 & 1.57 $\pm$ 0.07 E+2 & 1.78 $\pm$ 0.01 E+4 & 4.4 E+1 & 2.5 E+4 & 2.2 E+2 & 1.73 E+4 & 129 & 17 & 0.984 & 6.4\% & 2.1\% \\
20 & 4.43 $\pm$ 0.32 E+1 & 1.14 $\pm$ 0.01 E+4 & 8.3 E+0 & 1.6 E+4 & 6.0 E+1 & 1.12 E+4 & 44 & 14 & 0.959 & 9.6\% & 5.7\% \\
21 & 7.10 $\pm$ 0.29 E+1 & 1.51 $\pm$ 0.01 E+4 & 7.5 E+0 & 1.2 E+4 & 5.6 E+1 & 1.46 E+4 & 106 & 12 & 0.989 & 3.0\% & 9.8\% \\
22 & 1.67 $\pm$ 0.09 E+2 & 1.39 $\pm$ 0.01 E+4 & 6.6 E+1 & 2.1 E+4 & 2.6 E+2 & 1.34 E+4 & 135 & 15 & 0.977 & 1.9\% & 13\% \\
23 & 3.26 $\pm$ 0.14 E+1 & 1.23 $\pm$ 0.00 E+4 & 1.9 E+0 & 7.8 E+3 & 2.1 E+1 & 1.22 E+4 & 15 & 8.9 & 0.990 & 2.6\% & 22\% \\
\hline
avg. &  &  &  &  &  &  & 95 & 13 & 0.969 & 6.6\% & 12\% \\
s.d. &  &  &  &  &  &  & 78 & 5.2 & 0.022 & 5.5\% & 7.1\%
\enddata
\end{deluxetable}

\clearpage
\begin{deluxetable}{ccccccccc}
\tabletypesize{\small}
\tablecaption{Comparison of fitting effects between the 
modified logistic differential equation with modified Gaussian 
function and quasi-Plank function.
\label{tab:comparision}}
\tablehead{
\colhead{Sunspot} & \colhead{Function} & \colhead{Number of} & 
\colhead{$\overline{A^2}$} & \colhead{$\overline{\sigma}$} & 
\colhead{$\overline{r}$} & \colhead{$\overline{BIC}$} & 
\colhead{$\overline{\delta R_m /R_m}$} & 
\colhead{$\overline{\delta T_a /T_a}$} \\
\colhead{Version} & & \colhead{Parameters}
}
\startdata
\multirow{6}{1cm}{V1} & L\tablenotemark{a} & 4 & 24 & 5.8 & 0.984 & 851 & 5.3\% & 10\% \\
 & L & 3 & 39 & 6.4 & 0.980 & 874 & 5.5\% & 12\% \\
 & L & 2 & 59 & 8.0 & 0.969 & 917 & 6.8\% & 12\% \\
 & G\tablenotemark{b} & 4 & 53 & 5.2 & 0.986 & 823 & 5.2\% & 9.4\% \\
 & P\tablenotemark{c} & 3 & 25 & 6.8 & 0.978 & 885 & 7.7\% & 13\% \\
 \hline
\multirow{6}{1cm}{V2} & L & 4 & 39 & 9.3 & 0.984 & 981 & 5.2\% & 10\% \\
 & L & 3 & 64 & 10 & 0.980 & 1004 & 5.3\% & 12\% \\
 & L & 2 & 95 & 13 & 0.969 & 1048 & 6.6\% & 12\% \\
 & G & 4 & 89 & 8.4 & 0.986 & 953 & 5.1\% & 9.5\% \\
 & P & 3 & 42 & 11 & 0.978 & 1014 & 7.6\% & 13\%
\enddata
\tablenotetext{a}{Modified logistic differential equation.}
\tablenotetext{b}{Modified Gaussian function \citep{Du2011}.}
\tablenotetext{c}{Quasi-Plank function \citep{HathawayEA1994}.}
\end{deluxetable}

\clearpage
\begin{deluxetable}{ccccccc}
\tablecaption{Shannon entropy of each phase, which are calculated by the 
monthly sunspot number, and cycle length for solar cycles $1-23$.
\label{tab:cycleFeatures}}
\tablehead{
\colhead{Cycle No.} & \colhead{$E_1$} & \colhead{$E_2$} & \colhead{$E_3$} &
 \colhead{$E_4$} & \colhead{$E_5$} & \colhead{$T_c$}
}
\startdata
1 & 5.0 & 5.9 & 5.9 & 6.4 & 5.2 & 11.3 \\
2 & 5.6 & 6.7 & 7.6 & 6.8 & 6.0 & 9.0 \\
3 & 5.8 & 7.4 & 6.5 & 6.7 & 4.7 & 9.3 \\
4 & 6.3 & 6.6 & 5.8 & 6.0 & 5.3 & 13.6 \\
5 & 5.1 & 5.3 & 5.5 & 4.8 & 4.4 & 11.9 \\
6 & 4.0 & 5.2 & 6.6 & 6.0 & 5.1 & 13.0 \\
7 & 5.4 & 6.0 & 6.1 & 6.9 & 5.6 & 10.7 \\
8 & 6.3 & 7.4 & 6.9 & 6.9 & 5.7 & 9.7 \\
9 & 5.6 & 7.2 & 6.9 & 6.4 & 5.4 & 12.4 \\
10 & 4.9 & 6.5 & 6.5 & 6.2 & 5.6 & 11.3 \\
11 & 6.2 & 7.2 & 6.4 & 6.0 & 4.7 & 11.8 \\
12 & 5.6 & 6.5 & 6.3 & 5.7 & 4.6 & 11.3 \\
13 & 5.9 & 6.6 & 6.1 & 6.1 & 4.5 & 11.8 \\
14 & 5.7 & 6.6 & 7.2 & 6.3 & 4.4 & 11.5 \\
15 & 5.4 & 6.3 & 7.2 & 6.2 & 5.3 & 10.0 \\
16 & 5.5 & 6.4 & 6.2 & 6.5 & 4.7 & 10.2 \\
17 & 5.1 & 6.7 & 7.1 & 6.7 & 6.2 & 10.4 \\
18 & 5.8 & 6.7 & 7.2 & 6.2 & 5.7 & 10.2 \\
19 & 5.6 & 7.0 & 6.7 & 6.0 & 5.4 & 10.5 \\
20 & 5.2 & 6.5 & 5.5 & 5.9 & 5.3 & 11.4 \\
21 & 5.7 & 6.9 & 7.0 & 6.5 & 5.1 & 10.5 \\
22 & 5.8 & 6.7 & 6.7 & 6.2 & 5.2 & 9.7 \\
23 & 5.6 & 6.7 & 6.6 & 5.8 & 4.7 & 12.6
\enddata
\end{deluxetable}

\clearpage
\begin{deluxetable}{ccccccccc}
\tablecaption{Predicted parameters from TMLP model, the
 evaluating indices, and the absolute relative error 
 of cycle features for cycles $4-23$.
\label{tab:prediction}}
\tablehead{
\colhead{Cycle No.} & \colhead{$x_0$} & \colhead{$x_m$} &
 \colhead{$A^2$} & \colhead{$\sigma$} & \colhead{$CC$} &
 \colhead{$\frac{\delta S_m}{S_m}$} & \colhead{$\frac{\delta T_a}{T_a}$} &
 \colhead{$\frac{\delta T_c}{T_c}$}
}
\startdata
4 & 1.13 E+2 & 1.60 E+4 & 2.2 E+2 & 21 & 0.957 & 2.0\% & 15\% & 22\% \\
5 & 3.75 E+1 & 5.53 E+3 & 4.5 E+2 & 18 & 0.768 & 1.1\% & 71\% & 16\% \\
6 & 1.09 E+0 & 5.97 E+3 & 1.2 E+2 & 14 & 0.821 & 10\% & 5.7\% & 8.8\% \\
7 & 1.08 E+0 & 7.63 E+3 & 3.1 E+2 & 13 & 0.944 & 3.9\% & 13\% & 14\% \\
8 & 8.41 E+1 & 1.54 E+4 & 3.4 E+2 & 30 & 0.909 & 5.5\% & 18\% & 11\% \\
9 & 1.29 E+2 & 1.42 E+4 & 2.1 E+3 & 50 & 0.574 & 2.9\% & 20\% & 16\% \\
10 & 3.88 E+1 & 1.38 E+4 & 4.5 E+1 & 19 & 0.940 & 11\% & 7.4\% & 1.9\% \\
11 & 6.72 E+1 & 1.43 E+4 & 2.5 E+2 & 22 & 0.954 & 8.4\% & 18\% & 8.3\% \\
12 & 2.40 E+1 & 8.21 E+3 & 1.7 E+2 & 14 & 0.938 & 1.0\% & 11\% & 4.6\% \\
13 & 5.88 E+1 & 8.48 E+3 & 7.3 E+1 & 14 & 0.952 & 13\% & 4.2\% & 13\% \\
14 & 3.02 E+1 & 7.10 E+3 & 2.7 E+1 & 10 & 0.961 & 0.50\% & 3.9\% & 9.3\% \\
15 & 1.55 E+1 & 9.15 E+3 & 5.4 E+2 & 22 & 0.906 & 22\% & 14\% & 11\% \\
16 & 6.62 E+1 & 1.01 E+4 & 4.5 E+1 & 15 & 0.937 & 16\% & 19\% & 2.3\% \\
17 & 3.80 E+1 & 1.17 E+4 & 4.8 E+1 & 12 & 0.982 & 12\% & 19\% & 4.4\% \\
18 & 9.12 E+1 & 1.34 E+4 & 8.3 E+1 & 13 & 0.983 & 8.1\% & 19\% & 3.5\% \\
19 & 3.21 E+1 & 1.54 E+4 & 1.3 E+3 & 50 & 0.849 & 19\% & 16\% & 7.1\% \\
20 & 1.04 E+2 & 1.20 E+4 & 3.7 E+2 & 21 & 0.906 & 15\% & 6.5\% & 9.5\% \\
21 & 1.31 E+2 & 1.39 E+4 & 2.1 E+2 & 24 & 0.953 & 10\% & 2.2\% & 1.4\% \\
22 & 9.60 E+1 & 1.34 E+4 & 3.3 E+2 & 21 & 0.956 & 5.1\% & 21\% & 8.6\% \\
23 & 7.80 E+1 & 1.31 E+4 & 2.0 E+2 & 19 & 0.953 & 9.3\% & 35\% & 16\% \\
\hline
avg. &  &  & 3.6 E+2 & 21 & 0.907 & 8.8\% & 17\% & 9.5\% \\
s.d. &  &  & 4.9 E+2 & 11 & 0.095 & 6.2\% & 15\% & 5.6\%
\enddata
\end{deluxetable}

\clearpage
\begin{deluxetable}{ccccc}
\tablecaption{Comparison of prediction results between
TMLP with modified Gaussian function and quasi-Plank function.
\label{tab:prediction_v1}}
\tablehead{
\colhead{Function} & \colhead{$\Delta t$ (month)} & \colhead{$\delta R_m/R_m$}
 & \colhead{$\delta a/a$\tablenotemark{d}} & \colhead{$\delta T_c /T_c$}
}
\startdata
L\tablenotemark{a} & 15 & 12\% & --- & 9.2\% \\
G\tablenotemark{b} & 25 & 15\% & --- & --- \\
P\tablenotemark{c}    & 30 & --- & 20\% & --- \\
P    & 42 & --- & 10\% & ---
\enddata
\tablenotetext{a}{Modified logistic differential equation.}
\tablenotetext{b}{Modified Gaussian function \citep{Du2011}.}
\tablenotetext{c}{Quasi-Plank function \citep{HathawayEA1994}.}
\tablenotetext{d}{Parameter $a$ is the amplitude in Equation~(\ref{eq:quasi_Planck}).}
\end{deluxetable}

\end{document}